\documentclass{mn2e}
\usepackage{color}
\usepackage{hyperref}
\usepackage{graphicx}
\let\bibhang\undefined

\usepackage{natbib}
\usepackage{astroj}
\usepackage{lscape}
\usepackage[twoside]{rotating}
\usepackage{enumerate}
\usepackage{lettrine}
\usepackage{times}
\title[Quantifying Chemical Tagging]{Quantifying Chemical Tagging: Towards Robust Group Finding in the Galaxy}
\author[Mitschang et al.]{A. W. Mitschang$^{1,2}$ \thanks{E-mail: arik.mitschang@mq.edu.au}, G. De Silva$^{3}$, S. Sharma$^{4}$, D. B. Zucker$^{1,2,3}$\\
$^{1}$ Macquarie University Research Centre in Astronomy, Astrophysics \& Astrophotonics, NSW 2109, Australia\\
$^{2}$ Department of Physics \& Astronomy, Macquarie University, NSW 2109, Australia\\
$^{3}$ Australian Astronomical Observatory, PO Box 915, North Ryde, NSW 1670, Australia\\
$^{4}$ Sydney Institute for Astronomy, School of Physics, University of Sydney, NSW 2006, Australia\\
}
\begin{document}
\maketitle

\begin{abstract}
The first generation of large-scale chemical tagging surveys, in
particular the HERMES/GALAH million star survey, promises to vastly
expand our understanding of the chemical and dynamical evolution of
the Galaxy. This, however, is contingent on our ability to confidently
perform chemical tagging on such a large data-set. Chemical
homogeneity has been observed across a range of elements within
several Galactic open clusters, yet the level to which this is the
case globally, and particularly in comparison to the scatter across
clusters themselves, is not well understood. The patterns of elements
in coeval cluster members, occupying a complex chemical abundance
space, are rooted in the evolution, ultimately the nature of the very
late stages, of early generations of stars. The current astrophysical
models of such stages are not yet sufficient to explain all
observations, combining with our significant gaps in the understanding
of star formation, makes this a difficult arena to tackle
theoretically. Here, we describe a robust pair-wise metric used to
gauge the chemical difference between two stellar components. This
metric is then applied to a database of high-resolution literature
abundance sources to derive a function describing the probability that
two stars are of common evolutionary origin. With this cluster
probability function, it will be possible to report a confidence,
grounded in empirical observational evidence, with which clusters are
detected, independent of the group finding methods. This formulation
is also used to probe the role of chemical dimensionality, and that of
individual chemical species, on the ability of chemical tagging to
differentiate coeval groups of stars.
\end{abstract}
\section{Introduction}
It has long been postulated that stars form in dense clusters out of
an overdensity of primordial gas and dust (e.g. see
\citealt{shu1987}). Evidence for this is seen in present day open clusters,
but their numbers by no means account for the large stellar population
of the Galactic disk. \citet{janes1988} suggest that a typical lifetime
of an open cluster in the disk is 100 Myr, after which their stellar
components dissipate through dynamical interactions, explaining the
fact that stars are not strictly observed in clusters. It is possible
this dissipative action is seen in progress in so-called ``moving
groups'' and young associations
(e.g. \citealt{eggen1994,montes2001,zuckerman2004,torres2008}). A small
but significant population of very old open clusters have since been
identified \citep{phelps1994,janes1994,friel1995}, notably including
NGC 6791 with an age of $\sim${}7.2 Gyr \citep{kaluzny1995}, with the
zone in-between filled by a wide range of ages \citep{Xin2005}. Present
day star formation within embedded molecular clouds, such as the Orion
nebula, completes the circle of cluster evolution
\citep{lada2003}. These observations indicate that stars indeed have
formed and continue to form in clusters through all epochs of Galactic
evolution.

Although the particulars of star formation in molecular clouds remain
uncertain, the observed abundances of elements point to a successive
chain of star formation, chemical enrichment of the interstellar
medium, and subsequent formation events in a cloud, occurring over
generations of stars. Most of the elements known to exist in the
universe are processed in the very late stages or death throes of a
star (e.g. see \citealt{wallerstein1997}). The origin of the Fe-peak
elements is believed to be primarily in type Ia supernovae, while the
$\alpha${}-, and r-process elements are most likely predominantly
generated within the violent explosions of core-collapse (type II)
supernovae \citep{kratz2007}. The relatively more quiescent asymptotic
giant branch (AGB) phase, in low- to intermediate-mass stars, is
thought to be responsible for most of the s-processing
\citep{karakas2007,busso1999}, and thus is the sole origin of the pure
s-process elements and a contributor of elements produced by both the
s- and r-processes. The next generation of low mass stars, born out of
this freshly enriched gas and dust cloud, will spend a significant
amount of time on the main sequence and giant branches. In these
stages, and at these masses, the nuclear processing and convection
fueling the star are not sufficient to alter the photospheric
abundances of those elements \citep{iben1967}. Therefore, it is these
groups of elements that form the basis of the technique that has come
to be known as chemical tagging.

The technique of chemical tagging, first proposed by \citet{fbh2002},
promises to identify clusters of stars, even after being scattered by
the dissipative evolution of the disk, through patterns of abundances
in these key elements. To date, several important test have been made,
confirming that members of individual open clusters are chemically
homogeneous \citep{desilva2009,desilva2007b}, as well as less obvious
kinematically linked stellar aggregates known as moving groups
\citep{desilva2007a,bubar2010}. The technique has not, to our
knowledge, been applied on any scale to a random sample of stars, but
has been utilized to confirm or reject suspected membership amongst
kinematic groups in several recent studies
(e.g. \citealt{tabernero2012,pompeia2011,carretta2012,desilva2011}). A
particular challenge is the large number of dimensions that are
required in order to guarantee uniqueness across populations;
\citet{bhf2004} and \citet{desilva2007b} estimate between 10 and 15
elements, are necessary, given the number of formation sites and some
assumptions of abundance accuracy. These elements form the dimensions
of the so-called $\mathcal C$-space \citep{fbh2002}. The sample size of
studies with sufficient abundances, however, remains small.

The ambitious chemical tagging project known as the GALactic
Archaeology with HERMES
(GALAH\footnote{http://www.aao.gov.au/HERMES/GALAH/Home.html}) survey, to
be undertaken with the upcoming High Efficiency and Resolution
Multi-Element Spectrograph (HERMES;\citealt{barden2010}), will target a
million disk stars at high-resolution in a relatively short
time-span. The possibility of piecing back together some of the
presumably 10$^{8}$ relic clusters dissolved in the Galaxy \citep{bhf2004},
could vastly expand our knowledge of the star formation history and
structural and chemical evolution of the Galactic disk. The Gaia-ESO
Public Spectroscopic Survey \citep{gilmore2012}, an on-going chemical
abundance analysis survey, will target approximately 100000 field and
open cluster stars in the Galaxy at high resolution, providing a rich
dataset complementing GALAH. A very interesting prospect is the
search for Solar siblings, or the birth cluster of the Sun, and its
origin and possible journey within the Galaxy
\citep{bland-hawthorn2010a,batista2012}. The ability to tackle
these arenas is dependent on the ability to reliably distinguish
compatible from incompatible chemistries (i.e. between two stars in
such a survey).

We present here, for the first time, a metric that quantifies, in a
consistent yet flexible manner, global abundance trends in terms of
chemical tagging and show how a given population can be probed using
this metric, along with a representative cluster sample, without the
need to define an arbitrary ``level'' of chemical homogeneity. We show
how any stellar population can be chemically tagged, and how a
quantitative level of confidence can be applied to its members to
gauge the overall quality of a detection based solely on elemental
abundances. We further discuss some implications of this approach to
carrying out large scale chemical tagging surveys such as GALAH.
\section{A Metric to probe the chemical nature of clusters}
On the subject of chemical tagging, there has been little discussion
in the literature thus far on establishing \emph{quantitative} differences
in chemical signatures between different clusters and the stars within
a given cluster. Several authors have obtained and analyzed high
resolution data-sets on known cluster or moving group populations with
the aim of testing chemical tagging
\citep{desilva2007b,desilva2007a,desilva2009,bubar2010} with quite
promising results. These studies have taken a qualitative approach to
differences in chemical abundance patters. There has also been work
recently on understanding the complexity and interrelation of the
chemical abundance space using Principle Component Analysis
\citep{ting2012}. The goal there is to reduce the overall
dimensionality to only the strongest variant dimensions. That study,
however, was restricted to probing the chemical patterns \emph{between}
different populations, including open clusters, looking for the
largest scatter in nucleosynthesis yields. There is still lacking a
global characterization of the differences in patterns in $\mathcal
C$-space, e.g. the level of homogeneity expected within a cluster is
not clearly defined.

The difficulty in conceptualizing the interplay between all of the
dimensions of $\mathcal C$-space, combined with the non-uniformity of
the available abundance data, leads us to seek a description of the
chemical difference between clusters that is both independent of the
dimensionality of $\mathcal C$-space, and itself one-dimensional.
\subsection{A metric for determining chemical difference}
We define a simple metric for quantifying the chemical difference
between any two stars

\[
\delta_C=\sum_{C}^{N_C}\omega_C\frac{|A_{C}^i-A_{C}^j|}{N_C}
\]
Where $C$ is a particular chemical species out of the $N_C$ species
for which abundances have been determined, A$_{C}$ is the abundance ratio
of element $C$ to Fe relative to solar (i.e. [X/Fe] in standard
notation), except when $C$ is Fe, in which case it is the ratio of Fe
to H; $i$ and $j$ are the two stars for which this metric is being
computed; and $\omega_{C}$ represents a weighting factor for an individual
species. For this work, $\omega_{C}$ is always equal to one, but is
included in this definition for future investigations, when
sufficiently large data-sets, which will allow its examination, become
available. Simply put, $\delta_{C}$ is the mean absolute difference of
species $C$ between two stars across all $N_C$ chemical species. There
are two reasons for the choice of this type of metric (a Manhattan
type) as opposed to a Euclidean type. First and foremost, its use
facilitates the separation of clusters in $\mathcal C$-space, the
ultimate goal of chemical tagging, and secondly, the value of $\delta_{C}$
is more easily interpreted in the context of abundance
measurements. Another advantageous property of this metric definition,
due to the N$_{C}$ denominator, is that when considering any number of
correlated elements in the summation, the value of $\delta_{C}$ is
independent of the number of number of elements, $N_C$.

In this paper, the binned probability density distribution of $\delta_{C}$
for a single population, normalized to its maximum per-bin $\delta_{C}$
number density, shall be referred to as the $\eta_{C}$ distribution. When
such a population contains only members of a cluster, i.e. a single
group of coeval stars, chemical tagging theory posits that this
distribution should occupy the low end of the difference spectrum,
while populations derived from strictly distinct clusters will occupy
the high end. The following two terms describe these different
populations and will be used throughout this text. Their definitions
and uses should be understood as follows:

\begin{itemize}
\item  \textbf{intra-cluster}: All members of this group are confirmed members of
  a single open cluster. $\eta_{C}$ will be built from those $\delta_{C}$
  computed between any two stars that satisfy the above definition and
  for which any stated conditions apply (e.g. minimum number of
  elements, etc.)

\item  \textbf{inter-cluster}: Members of this group are confirmed members of any
  open cluster. $\eta_{C}$ will be built from all those $\delta_{C}$ computed
  between two stars that are \emph{not} members of the same open cluster
  and for which any stated conditions apply.

\end{itemize}

\begin{figure}
\centering
\includegraphics[width=1\linewidth]{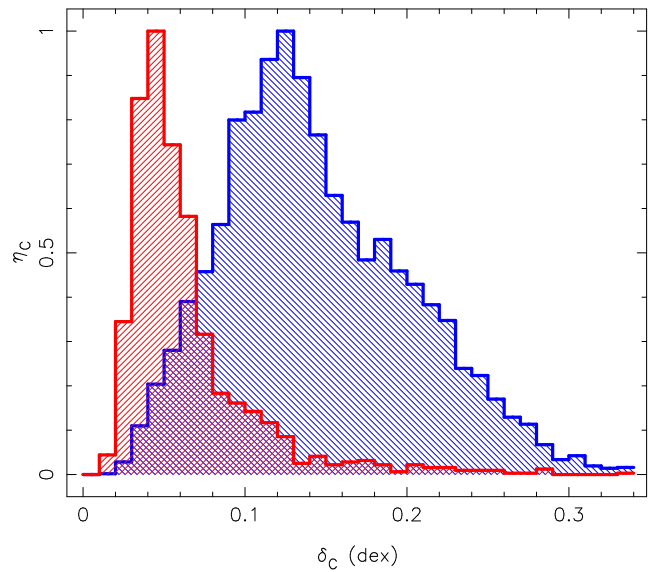} 
\caption{The $\eta_{C}$ distributions for intra- (red; left) and inter-cluster
(blue; right) populations. The total range of the inter-cluster distribution
is larger than shown, and has been cut off as described in the text.
 \label{deltacfig}}
\end{figure}

Figure \ref{deltacfig} shows both intra- and inter-cluster $\eta_{C}$
distributions for the clusters listed in Table \ref{deltactab},
calculated with the condition that a minimum of 8 elements be used in
the $\delta_{C}$ computation (for a detailed explanation of this choice,
see section \ref{sec:explore}). We note that the total range of the
inter-cluster distribution is much larger than that depicted, however
it is not instructive to display anything beyond the range of the
intra-cluster distribution for the purposes of our analysis. It is
important also to note that, because we do not assume a cluster is
represented by its mean abundances, the inter-cluster distribution
takes into account all scatter within a given cluster, across all
chemical dimensions. This point is crucial to our analysis, and its
potential usefulness in terms of a true chemical tagging experiment,
in that we wish to determine the ability to distinguish individual
pairs of stars. Assuming perfect homogeneity amongst cluster members
in this respect would likewise circumvent our treatment of the
intra-cluster distribution.

\subsection{Open cluster abundance data}
Table \ref{deltactab} gives a summary of the literature elemental
abundance database generated for use in this analysis. Each row in the
table is a particular study, from the literature, of a single cluster
as indicated. Basic information includes the cluster name, the number
of stars for which abundances were derived ($N_\star$), the total number
of element species derived ($N_C$), and the literature reference. The
remaining columns in the table summarize the abundance data: for each
element existing in the database, the number of stars for which
abundances for that element were derived is listed. The numbers in
this matrix are equal to or less than the corresponding $N_\star$,
depending on, for example, the range or quality of a particular
spectrum.

\begin{table*}
\rotatebox{90}{
\begin{minipage}{\textheight}
\begin{center}
\caption{Summary of the high-resolution literature abundance database
 \label{deltactab}}
\begin{tabular}{llllllllllllllllllllllllllll}
\hline
 Cluster   &  Source$^{a}$  &  $N_\star$  &  $N_C$  &  Na  &  Mg  &  Al  &  Si  &  Ca  &  Sc  &  Ti  &  V   &  Cr  &  Mn  &  Fe  &  Co  &  Ni  &  Cu  &  Zn  &  Sr  &  Y   &  Zr  &  Ba  &  La  &  Ce  &  Nd  &  Sm  &  Eu \\
\hline
 Pleiades  &  A         &         20  &      5  &  20  &   -  &  -   &  20  &   -  &  -   &  20  &  -   &   -  &  -   &  20  &  -   &  20  &  -   &  -   &  -   &  -   &  -   &   -  &  -   &  -   &  -   &  -   &  -  \\
 IC2391    &  B         &          7  &      6  &   7  &   -  &  -   &   7  &   7  &  -   &   7  &  -   &   -  &  -   &   7  &  -   &   7  &  -   &  -   &  -   &  -   &  -   &   -  &  -   &  -   &  -   &  -   &  -  \\
 IC2602    &  B         &          8  &      6  &   8  &   -  &  -   &   8  &   8  &  -   &   8  &  -   &   -  &  -   &   8  &  -   &   8  &  -   &  -   &  -   &  -   &  -   &   -  &  -   &  -   &  -   &  -   &  -  \\
 Blanco1   &  C         &          8  &      6  &   -  &   8  &  -   &   8  &   8  &  -   &   8  &  -   &   -  &  -   &   8  &  -   &   8  &  -   &  -   &  -   &  -   &  -   &   -  &  -   &  -   &  -   &  -   &  -  \\
 Be29      &  D         &          6  &      6  &   -  &   -  &  -   &   6  &   6  &  -   &   6  &  -   &   6  &  -   &   -  &  -   &   6  &  -   &  -   &  -   &  -   &  -   &   6  &  -   &  -   &  -   &  -   &  -  \\
 Cr261     &  D         &          7  &      6  &   -  &   -  &  -   &   7  &   7  &  -   &   7  &  -   &   7  &  -   &   -  &  -   &   7  &  -   &  -   &  -   &  -   &  -   &   7  &  -   &  -   &  -   &  -   &  -  \\
 Mel66     &  D         &          6  &      6  &   -  &   -  &  -   &   6  &   6  &  -   &   6  &  -   &   6  &  -   &   -  &  -   &   6  &  -   &  -   &  -   &  -   &  -   &   6  &  -   &  -   &  -   &  -   &  -  \\
 IC4665    &  E         &         18  &      6  &   -  &  18  &  -   &  18  &  18  &  -   &  18  &  -   &  18  &  -   &   -  &  -   &  18  &  -   &  -   &  -   &  -   &  -   &   -  &  -   &  -   &  -   &  -   &  -  \\
 Be32      &  F         &          9  &      8  &   9  &   -  &  -   &   9  &   9  &  -   &   9  &  -   &   9  &  -   &   9  &  -   &   9  &  -   &  -   &  -   &  -   &  -   &   9  &  -   &  -   &  -   &  -   &  -  \\
 NGC2324   &  F         &          7  &      8  &   7  &   -  &  -   &   7  &   7  &  -   &   7  &  -   &   7  &  -   &   7  &  -   &   7  &  -   &  -   &  -   &  -   &  -   &   7  &  -   &  -   &  -   &  -   &  -  \\
 NGC2477   &  F         &          6  &      8  &   6  &   -  &  -   &   6  &   6  &  -   &   6  &  -   &   6  &  -   &   6  &  -   &   6  &  -   &  -   &  -   &  -   &  -   &   6  &  -   &  -   &  -   &  -   &  -  \\
 NGC2660   &  F         &          5  &      8  &   5  &   -  &  -   &   5  &   5  &  -   &   5  &  -   &   5  &  -   &   5  &  -   &   5  &  -   &  -   &  -   &  -   &  -   &   5  &  -   &  -   &  -   &  -   &  -  \\
 NGC3960   &  F         &          6  &      8  &   6  &   -  &  -   &   6  &   6  &  -   &   6  &  -   &   6  &  -   &   6  &  -   &   6  &  -   &  -   &  -   &  -   &  -   &   6  &  -   &  -   &  -   &  -   &  -  \\
 IC4651    &  G         &          5  &      8  &   5  &   -  &  5   &   5  &   5  &  -   &   5  &  -   &   5  &  -   &   5  &  -   &   5  &  -   &  -   &  -   &  -   &  -   &   -  &  -   &  -   &  -   &  -   &  -  \\
 M67       &  G         &          6  &      8  &   6  &   -  &  4   &   6  &   6  &  -   &   6  &  -   &   6  &  -   &   6  &  -   &   6  &  -   &  -   &  -   &  -   &  -   &   -  &  -   &  -   &  -   &  -   &  -  \\
 Praesepe  &  G         &          7  &      8  &   7  &   -  &  5   &   7  &   7  &  -   &   7  &  -   &   7  &  -   &   7  &  -   &   7  &  -   &  -   &  -   &  -   &  -   &   -  &  -   &  -   &  -   &  -   &  -  \\
 NGC6192   &  H         &          4  &      8  &   -  &   4  &  4   &   4  &   4  &  -   &   4  &  -   &   4  &  -   &   4  &  -   &   4  &  -   &  -   &  -   &  -   &  -   &   -  &  -   &  -   &  -   &  -   &  -  \\
 NGC6404   &  H         &          4  &      8  &   -  &   4  &  4   &   4  &   4  &  -   &   4  &  -   &   4  &  -   &   4  &  -   &   4  &  -   &  -   &  -   &  -   &  -   &   -  &  -   &  -   &  -   &  -   &  -  \\
 Cr261     &  I         &         12  &      9  &  12  &  12  &  -   &  12  &  12  &  -   &   -  &  -   &   -  &  12  &  12  &  -   &  12  &  -   &  -   &  -   &  -   &  12  &  12  &  -   &  -   &  -   &  -   &  -  \\
 M67       &  J         &         10  &      9  &  10  &  10  &  10  &  10  &  10  &  -   &  10  &  -   &  10  &  -   &  10  &  -   &  10  &  -   &  -   &  -   &  -   &  -   &   -  &  -   &  -   &  -   &  -   &  -  \\
 NGC3532   &  M         &          5  &      9  &   -  &   5  &  -   &   5  &   5  &  5   &   5  &  5   &   5  &  -   &   -  &  5   &   5  &  -   &  -   &  -   &  -   &  -   &   -  &  -   &  -   &  -   &  -   &  -  \\
 NGC4756   &  M         &          5  &      9  &   -  &   5  &  -   &   5  &   5  &  5   &   5  &  5   &   5  &  -   &   -  &  5   &   5  &  -   &  -   &  -   &  -   &  -   &   -  &  -   &  -   &  -   &  -   &  -  \\
 NGC5822   &  M         &          5  &      9  &   -  &   3  &  -   &   5  &   5  &  5   &   5  &  5   &   5  &  -   &   -  &  5   &   5  &  -   &  -   &  -   &  -   &  -   &   -  &  -   &  -   &  -   &  -   &  -  \\
 NGC6791   &  K         &          4  &      9  &   -  &   4  &  4   &   4  &   4  &  4   &   4  &  -   &   -  &  4   &   -  &  -   &   4  &  -   &  -   &  -   &  -   &  -   &   4  &  -   &  -   &  -   &  -   &  -  \\
 M11       &  L         &          4  &     10  &   4  &   -  &  4   &   4  &   4  &  4   &   4  &  -   &   -  &  -   &   -  &  3   &   4  &  -   &  -   &  -   &  -   &  -   &   -  &  4   &  -   &  -   &  -   &  4  \\
 NGC6253   &  K         &          4  &     10  &   4  &   4  &  4   &   4  &   4  &  4   &   -  &  -   &   -  &  4   &   -  &  -   &   4  &  -   &  -   &  -   &  -   &  -   &   4  &  -   &  -   &  -   &  -   &  4  \\
 Hyades    &  N         &         46  &     12  &  46  &  46  &  -   &  46  &  46  &  -   &  46  &  -   &   -  &  -   &  46  &  -   &   -  &  -   &  46  &  -   &  -   &  46  &  46  &  18  &  46  &  46  &  -   &  -  \\
 Cr261     &  O         &          6  &     13  &   6  &   6  &  6   &   6  &   6  &  6   &   6  &  -   &   6  &  6   &   6  &  6   &   6  &  -   &  -   &  -   &  -   &  -   &   6  &  -   &  -   &  -   &  -   &  -  \\
 NGC3114   &  P         &          7  &     13  &   -  &   5  &  6   &   7  &   7  &  7   &   7  &  -   &   7  &  -   &   -  &  -   &   7  &  -   &  -   &  -   &  5   &  6   &   -  &  7   &  6   &  6   &  -   &  -  \\
 NGC3680   &  Q         &         11  &     14  &  11  &  11  &  8   &  11  &  11  &  -   &  11  &  -   &  11  &  -   &  11  &  -   &  11  &  -   &  11  &  -   &  11  &  -   &   8  &  7   &  -   &  8   &  -   &  -  \\
 Berenice  &  R         &         11  &     15  &  10  &  11  &  -   &   6  &   -  &  6   &  11  &  11  &  10  &  10  &  11  &  11  &  11  &  -   &  -   &  6   &  6   &  11  &  11  &  -   &  -   &  -   &  -   &  -  \\
 Pleiades  &  S         &          5  &     16  &   5  &   5  &  -   &   5  &   5  &  5   &   5  &  5   &   5  &  5   &   5  &  5   &   5  &  -   &  -   &  5   &  5   &  5   &   5  &  -   &  -   &  -   &  -   &  -  \\
 M67       &  T         &          9  &     20  &   9  &   9  &  9   &   9  &   9  &  9   &   9  &  9   &   9  &  9   &   9  &  9   &   9  &  9   &  -   &  -   &  9   &  9   &   9  &  9   &  9   &  -   &  -   &  9  \\
 NGC2360   &  U         &          4  &     21  &   4  &   4  &  4   &   4  &   4  &  4   &   4  &  4   &   4  &  3   &   4  &  4   &   4  &  4   &  2   &  -   &  4   &  4   &   3  &  3   &  2   &  -   &  3   &  -  \\
 NGC752    &  U         &          4  &     23  &   4  &   4  &  4   &   4  &   4  &  4   &   4  &  4   &   4  &  4   &   4  &  4   &   4  &  4   &  3   &  -   &  4   &  4   &   4  &  4   &  4   &  4   &  3   &  4  \\
\end{tabular}
\end{center}
\vskip 2.5mm {}$^{a}$ References using designations listed in Table \ref{sources}
\end{minipage}
}
\end{table*}

\begin{table}
\begin{center}
\caption{Sources and designations used in Table \ref{deltactab}
 \label{sources}}
\begin{tabular}{ll}
\hline
 Source reference              &  Label \\
\hline
 \citet{2009AJ....138.1292S}      &  A     \\
 \citet{2009AandA...501..553D}    &  B     \\
 \citet{2005MNRAS.364..272F}      &  C     \\
 \citet{2008AandA...488..943S}    &  D     \\
 \citet{2005ApJ...635..608S}      &  E     \\
 \citet{2008AandA...480...79B}    &  F     \\
 \citet{2008AandA...489..403P}    &  G     \\
 \citet{2010AandA...523A..11M}    &  H     \\
 \citet{desilva2007b}             &  I     \\
 \citet{2006AandA...450..557R}    &  J     \\
 \citet{2007AandA...473..129C}    &  K     \\
 \citet{2000PASP..112.1081G}      &  L     \\
 \citet{2009AandA...502..267S}    &  M     \\
 \citet{desilva2006,paulson2003}  &  N     \\
 \citet{2005AandA...441..131C}    &  O     \\
 \citet{2010IAUS..266..495P}      &  P     \\
 \citet{mitschang2012}            &  Q     \\
 \citet{2008AandA...479..189G}    &  R     \\
 \citet{2008AandA...483..567G}    &  S     \\
 \citet{2000AandA...360..499T}    &  T     \\
 \citet{2012MNRAS.419.1350}       &  U     \\
\end{tabular}
\end{center}
\end{table}

The table was generated first using sources listed in Table 12 of
\citet{carrera2011}, and then by a literature search for any other
studies based on high-resolution data for open clusters with
abundances of those elements listed in Table \ref{deltactab}. Due to the
sensitivity of computed abundances, especially when derived from
equivalent widths (EWs), with respect to resolution, and in order to
best match upcoming chemical tagging surveys, we only considered
studies whose typical resolution was better than R$\sim${}28,000. We
also only considered individual clusters which had at least four stars
for which abundances were derived. This latter condition is an
optimisation between the total number of clusters included and the
number of pairs used for the intra-cluster distribution, ensuring
adequate statistical significance for both populations. There proved
to be no reason to limit the number of chemical species derived in a
particular study ($N_C$); indeed the range allows us to probe some of
the effects of chemical dimensionality on the technique of chemical
tagging, which will be discussed in more detail in Section
\ref{sec:explore}.

All abundance values collected in the database are with respect to Fe
and compared to solar, i.e. [X/Fe] in standard notation. Fe abundances
are the only exception and are tabulated as [Fe/H]. Where a literature
study reported abundances otherwise, those values were converted to
this system when possible, and otherwise omitted. Care was taken to
ensure that the studies targeted unevolved stars. Uncertainties, where
available, are included in the database and are used to inform typical
values as discussed later in the text.

The final database with individual abundances, which is available as a
machine readable table with the online version of this text and on the
web\footnote{Note: To be uploaded to Vizier}, contains 35 individual
cluster studies of a total of 30 distinct clusters (clusters from
separate studies are listed separately in Table \ref{deltactab}) from 22
literature sources (those listed in Table \ref{sources}). There are a
total of 2775 individual abundance values measured for 291 stars, with
a minimum and maximum chemical dimensionality of 5 and 23,
respectively. A stub of the database is given in Table
\ref{abunddbstub}, where CID is a unique identifier for a given cluster
in a given study, and SID is a unique identifier for a given
star. These identifiers make it simple to disambiguate stars with
repeated names and clusters from different studies. The complete
version of Table \ref{abunddbstub} is available both online and in the
electronic version of this paper. The tabulated abundance values from
the literature were parsed via a computer program insofar as possible,
and by hand in some cases where the formats were not as uniform. In
very rare cases a literature study that matched our criteria may have
proved to be too difficult to parse and load (e.g. a scanned hard
copy) and had to be omitted.

\begin{table}
\begin{center}
\caption{Literature abundance database columns format $^{\dagger}$
 \label{abunddbstub}}
\begin{tabular}{lllllcrl}
\hline
 Cluster   &       CID  &  Star  &   SID  &  Src$^{a}$  &  El   &  [X/Fe]$^{b}$  &  Error \\
\hline
 NGC2360   &        50  &    12  &  1046  &  U      &  Al   &     -0.01  &   0.05 \\
 NGC2360   &        50  &    12  &  1046  &  U      &  Ca   &      0.13  &   0.06 \\
 NGC2360   &        50  &    12  &  1046  &  U      &  Co   &      0.03  &   0.09 \\
 NGC2360   &        50  &    12  &  1046  &  U      &  Cr   &      0.02  &   0.05 \\
 NGC2360   &        50  &    12  &  1046  &  U      &  Cu   &     -0.13  &        \\
 $\vdots$  &  $\ddots$  &        &        &         &       &            &        \\
\end{tabular}
\end{center}
\vskip 2.5mm $^{\dagger}$ The complete table with a detailed description of its
contents is available in the electronic version of this text.

$^a$ As in Table \ref{sources}

$^b$ Except for Fe $\rightarrow$ [Fe/H]
\end{table}

\subsection{The intra- and inter-cluster $\eta_{C}$ distributions}
Upon inspection of Figure \ref{deltacfig} it is evident that the two
populations have peaks significantly separated in $\delta_{C}$ space. Each
peak can be thought of as the ``characteristic'' chemical difference of
that population. In other words, based on Figure \ref{deltacfig}, we
find that Galactic open clusters, as represented by our literature
abundance database, typically differ chemically by $\sim${}0.120 dex,
while open clusters' \emph{internal} scatter is typically at the
$\sim${}0.045 dex level. Next we draw the readers attention to the
crossing point of the two distributions at $\sim${}0.07 dex. This
represents the point at which a pair of stars would be equally likely
to inhabit the same cluster as they would to be members of entirely
different clusters. Pairs of stars whose $\delta_{C}$ computes to the
right of this line are then more likely than not to be of distinct
origin, while those whose $\delta_{C}$ computes to the left would be
considered more likely to be siblings.

Another obvious feature is the bump in inter-cluster $\eta_{C}$ close to
0.19 dex. This is most likely due to noise on account of the limited
numbers of clusters in this work and would smoothe out if more data
were available. The extended tail toward high $\delta_{C}$ in the
intra-cluster distribution may also be attributed to noise, perhaps in
the form of mis-identified group members. Another possible explanation
for these ``outliers'' is related to errors in their abundance
measurements.  If not a noise feature, it may represent some form of
cloud fragmentation in the progenitor of those clusters, giving rise
to multi-modal star formation efficiencies within the same molecular
cloud. In any case, as we shall demonstrate, neither of these features
affects the probability analysis in the rest of this paper to any
significant degree.
\section{Systematics}
\label{sec:systematics}
Systematic uncertainties folded into abundance determinations come in
many forms. Among these are the methods used in computing EWs (and
those used to determine continuum levels); the choice of atmospheric
models including assumption of Local Thermodynamic Equilibrium
(LTE)/non-LTE calculations and convective overshooting; the synthetic
spectrum codes used; and the solar abundance values used to compute
relative abundances, to name just a few. In general, an individual
study will use identical techniques and parameters relating to these
factors in order to maintain internal consistency. As such, we do not
believe that systematic effects significantly alter the intra-cluster
distribution computed from our database from the ``natural'' one.

The story for \emph{inter}{}-cluster comparison is much more complicated.
Experience has shown that even with high quality high-resolution data,
the latest models and codes and very careful analysis, slight
differences in approach between studies can yield significantly
different results. Assuming an ``intrinsic'' inter-cluster $\eta_{C}$ with
some variance and mean $\delta_{C}$, introducing systematics will cause a
broadening and a shift in the mean. Searching for an estimate of the
intrinsic distribution, such that adding a systematic treatment will
result in an adequate representation of the data, is unrealistic given
the complexity of inter-cluster pairs in the database. Systematics,
for these purposes, should only affect pairs made from different
literature sources. The differing number of clusters and stars per
literature source also makes any simplified treatment impossible.  To
gauge the magnitude of both systematic effects, the mean shift and
broadening, for the specific case of our database, we induce
uncertainties in the data and extrapolate backward to a best guess of
the intrinsic distribution.

\begin{figure}
\centering
\includegraphics[width=1\linewidth]{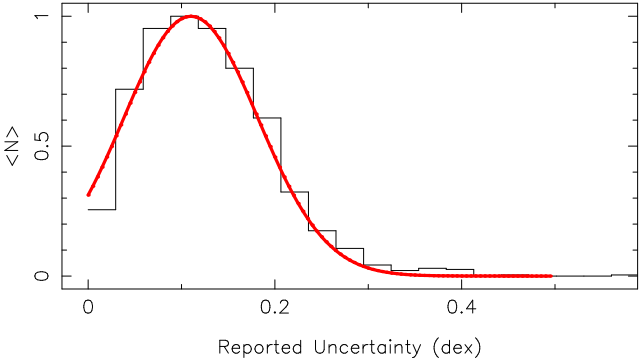} 
\caption{Gaussian fit (red) to the normalized distribution of
uncertainties reported in the literature (black histogram).
 \label{uncertainties}}
\end{figure}

To accomplish this a Monte-Carlo type of approach was taken, where in
each simulation iteration, we added randomly derived uncertainties to
the abundances of similar elements for all unique sources. In other
words, the same random value was added to all the measurements of a
given chemical species in a given literature study, regardless of
cluster membership. The absolute values of our randomly derived
uncertainties were taken from a Gaussian distribution whose parameters
fit the distribution of uncertainties recorded in the database. Figure
\ref{uncertainties} shows both the literature uncertainties and our
Gaussian model, which provides a reasonable fit to the data. The
inter-cluster $\eta_{C}$ is then computed from this set of simulated
cluster abundances and can be compared to the true inter-cluster
$\eta_{C}$ from our database.

\begin{figure*}
\centering
\includegraphics[width=1\linewidth]{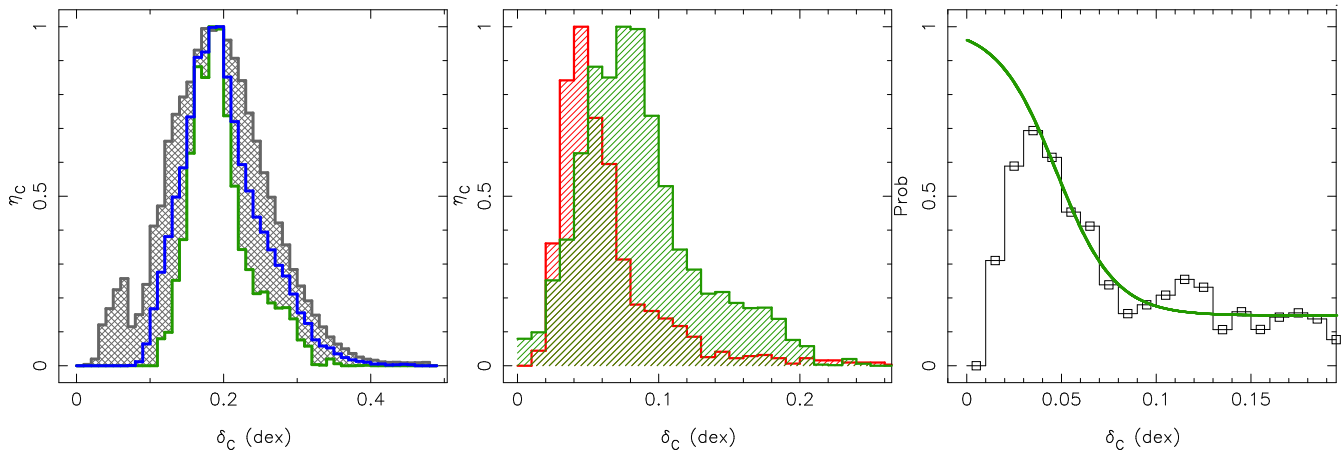} 
\caption{Summary of simulated inter-cluster systematics
treatment. The left panel shows the $\eta_{C}$ from 1000 systematics
simulations (grey line), along with the peak-shifted data (blue) and
the de-broadened and normalized ``intrinsic'' distribution
(green). The center panel shows the ``intrinsic'' distribution
peak-shifted to smaller $\delta_{C}$ based on the shift from our
simulations. The rightmost panel shows the probability function
(see Section \ref{sec:probfunc})
derived from the ``intrinsic'' $\eta_{C}$, and fit assuming the first two
bins are an artefact of this crude treatment.
 \label{sysp}}
\end{figure*}

We computed 1000 iterations of the systematics simulation described
above and derived $\eta_{C}$ across all simulations. The resultant
distribution is shown in the left panel of Figure \ref{sysp} (grey),
along with the data (blue), which have been shifted to align the
peaks. The broadening due to simulated abundance uncertainties is
evident. To reverse this process, or de-broaden, the difference
between simulation and data is removed from the data to arrive at the
green histogram in Figure \ref{sysp}, left panel. The grey
cross-hatching illustrates the total change between simulations, data
and the de-broadened distribution. Addition of the simulated
uncertainties caused the mean $\delta_{C}$ in the resultant distribution
to be shifted to a $\delta_{C}$ 56\% larger than that of the original
distribution. When corrected for the de-broadened distribution, in
order to arrive at an estimate of the intrinsic distribution, we find
the $\delta_{C}$ for which a 56\% increase arrives at the data mean, this
results in a peak at $\delta_{C}$ of $\sim${}0.077 dex shown in the center
panel of Figure \ref{sysp}, which is a repeat of Figure \ref{deltacfig}
using the ``intrinsic'' inter-cluster $\eta_{C}$ instead of the data.

This is perhaps a crude approximation, but should nonetheless give an
idea of the magnitude of potential effects. The uncertainties used,
for example, are internal random errors computed in a variety of ways
and may not represent the sources of error between studies as
well. Specific considerations, however, are beyond the scope of this
work. The bulk shift of our de-broadened distribution likely
mis-represents the true rising edge. Specifically, the first two,
perhaps three, bins shown in the center panel of Figure \ref{sysp} are
probably artefacts of this. Another artefact is visible in the
simulated $\eta_{C}$ in the left panel of Figure \ref{sysp} at a $\delta_{C}$ of
$\sim${}0.05 dex. The origin of this feature is unclear, but, given its
location, it does not affect our analysis in any meaningful way.

Internal, random, errors in abundance measurements will also alter the
$\eta_{C}$ distributions from their intrinsic profiles. Depending on the
overlap between clusters in $\mathcal C$-space, the magnitude of this
effect will be different for either population; however, the scatter
induced in the intra-cluster population requires that its magnitude be
greater. This implies that the overlap between $\eta_{C}$ distributions
seen in Figure \ref{sysp}, is an absolute upper limit, and the peaks
would be separated by a greater distance in $\delta_{C}$, favoring the
distinction of clusters. The specific impact that systematic effects,
as discussed here, have on our conclusions in this work will be
discussed in further detail in the next section.

\section{The cluster probability function}
\label{sec:probfunc}
\begin{figure}
\centering
\includegraphics[width=1\linewidth]{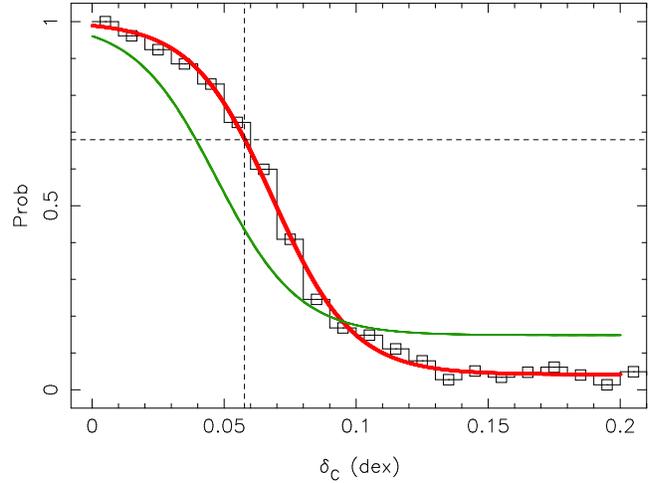} 
\caption{The intra-cluster probability function. The black squares
and histogram show the discrete probability function for the same
$\eta_{C}$ distributions shown in Figure \ref{deltacfig}. The solid red
line shows the least squares fit to this using the continuous
function with parameters listed in Table \ref{fitparams}, where the
dashed cross-hairs represent the 68\% confidence limit that a pair of
stars are coeval. Shown in green is the continuous fit for the
``intrinsic'' inter-cluster $\eta_{C}$ derived from our systematics
considerations.
 \label{prob8fig}}
\end{figure}

A major goal of this work is to derive a quantitative assessment of the
ability to distinguish coeval stars from those born in distinct
environments, based solely on their chemical signatures. The level of
homogeneity that is expected within a Galactic open cluster, and the
sparseness of the chemical space between them, are readily inferred
from Figure \ref{deltacfig}. We now combine these two concepts to
achieve this goal.  The probability that a particular pair of stars
are members of the same cluster based on their $\delta_{C}$ is related to
the relative contributions of each $\eta_{C}$ distribution, and can be
expressed as

\[
P_{\delta_C} \equiv prob(\delta_C) = \frac{\eta_{C}^{intra}}{\eta_{C}^{inter} + \eta_{C}^{intra}}
\]
When computed for the specific $\eta_{C}$ distributions plotted in Figure
\ref{deltacfig}, we obtain the discrete probability function shown in
Figure \ref{prob8fig}, where the cross-hairs represent the 68\%
confidence limit. Note the steepness of the function; by the time it
reaches a $\delta_{C}$ of 0.1 dex, there is only approximately a 10-20\%
probability that those two stars are members of the same cluster. This
places tight constraints on the performance of a chemical tagging
survey (e.g. see Section \ref{sec:reconstruction}).

\subsection{A functional form of $P_{\delta_C}$}
\label{sec:contprobfunc}
We now seek a functional form to express the intra-cluster probability
continuously as a function of $\delta_{C}$. A modified form of the
sigmoid function with four free parameters describes the data well and
can be written as

\[
prob(\delta_C)=c+\frac{1-c}{1+ae^{b\delta_C-d}}
\]
The $c$ term accounts for the lower probability limit caused by the
tail in the intra-cluster distribution, while the $1-c$ term ensures
that we do not obtain a probability greater than 1. The solid red line
in Figure \ref{prob8fig} shows the least squares best fit to the
discrete probability function. The goodness of fit illustrates both
the smooth nature of the $\eta_{C}$ distributions, and the appropriateness
of the continuous probability function.

Also shown in Figure \ref{prob8fig} is a fit to the probability function
from our derived ``intrinsic'' inter-cluster $\eta_{C}$ (see Section
\ref{sec:systematics}). This fit (shown in the right panel of Figure
\ref{sysp}, along with its discrete counterpart) assumed that the first
bin, corresponding to a $\delta_{C}$ of zero, or identical chemistries,
actually approaches one. Likewise, the second bin was ignored, both of
these under the assumption that they are noise features arising from
our crude systematics treatment.

The magnitude of possible systematic effects is quite visible. At the
68\% confidence level the difference between the two functions is
$\sim${}0.015 dex. It is difficult to judge whether our ``intrinsic''
function actually represents such, but Figure \ref{prob8fig} does help
to illustrate the advantages of homogeneous abundance analysis, even
if only to avoid the need for such considerations as applied here.
The four parameters corresponding to each fit are listed in Table
\ref{fitparams}.

\begin{table}
\begin{center}
\caption{Best fit parameters for $P_{\delta_C}$
 \label{fitparams}}
\begin{tabular}{ccc}
\hline
 Parameter  &  Data   &  Systematics simulation \\
\hline
         a  &  13.37  &  9.164                  \\
         b  &  65.51  &  64.41                  \\
         c  &  0.041  &  0.148                  \\
         d  &  7.060  &  5.242                  \\
\end{tabular}
\end{center}
\end{table}

This formulation now provides us with a way of not only filtering the
results of cluster finding algorithms, but importantly, a means to
report the confidence with which we detect them. There will be a rate
of attrition due to the overlap in the respective distributions, an
effect which is likely due to natural scatter in addition to complex
and not well understood processes occurring in the progenitors of all
open clusters. We stress the importance of arriving at this
probability function through empirical analysis of observational
data. Until we can be confident that we fully understand all of the
intricacies of star formation and Galactic chemical evolution, this
empirical approach will remain the \emph{only} way to achieve this.

This is not to say that the parameters listed in Table \ref{fitparams}
are not subject to change. These most likely represent a lower limit
to the ability to chemically tag, due to the heterogeneous analysis
methods present in our abundance database. In the case of a large
scale chemical abundance survey such as GALAH, or Gaia-ESO, a
probability function specific to the analysis methods of that survey
would be derived - so long as an appropriate calibration program
targeting a number of established coeval populations were adopted. The
analysis described in the present paper would be repeated using
cluster abundance data from that calibration program. The Gaia-ESO
survey serendipitously has this calibration built into its design,
since it aims to target 100 Galactic open clusters. Possibly, though
not ideally, this could serve as calibration for GALAH. For
heterogeneous chemical tagging experiments (without the benefits of a
uniform abundance analysis), the most up-to-date abundance data would
be added to the literature abundance database and a new probability
function derived. In such a way, the function parameters could be
iteratively improved. In the meantime, the function with parameters
listed in Table \ref{fitparams} represent our current best guess, within
the limitations of the abundance data in our database.
\section{Explorations in $\mathcal C$-space}
\label{sec:explore}
Up to this point, all our discussion has been involving the $\eta_{C}$
distributions whose $\delta_{C}$ metric computations were required to have
at least a dimensionality of eight (i.e., a minimum of eight distinct
elements being compared). This is not an entirely arbitrary decision,
but one that is based both on the practicality of future chemical
tagging experiments, and a delicate balance between optimising the
cluster probability function and maintaining the statistical
significance of the distributions themselves.

\begin{figure}
\centering
\includegraphics[width=1\linewidth]{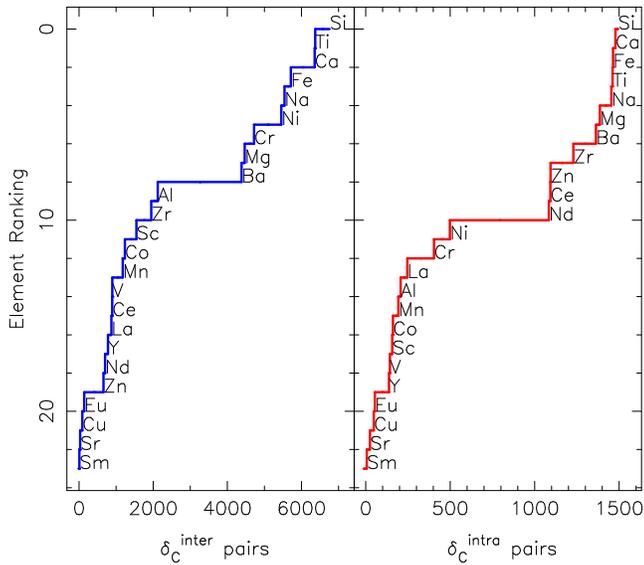} 
\caption{The numbers of $\delta_{C}$ computations making up each
distribution shown in Figure \ref{deltacfig}. The ``ranking''
increases from top to bottom, where a lower number rank indicates
greater significance in the overall distribution. This is \emph{not} a
reflection of the significance of any individual element as a
chemical tagging probe, but rather informs the groups of elements
generally necessary to utilise the cluster probability function
described here.
 \label{deltacchar8}}
\end{figure}

\citet{bhf2004} estimated the minimum number of dimensions required to
reveal a fraction of the approximately 10$^{8}$ clusters believed to be
dissolved in the Galaxy. This figure, between 10-15 elements, has been
incorporated into the design of the HERMES instrument for the purpose
of the GALAH chemical tagging survey. \citet{ting2012} confirm this
number finding approximately 9 independently varying elements from the
available abundance data. The constraint employed here matches well
these capabilities, in fact, the mean number of elements in the intra-
and inter-cluster $\eta_{C}$ distributions came out to be 11 and 9,
respectively. In other words, out of the total number of $\delta_{C}$
pairs in each distribution, half were computed with greater than, or
equal to, 11 and 9 compatible element abundances. It is important,
also, to understand the species that dominate each of the
distributions. These will be the minimum set of species required to
utilize the cluster probability function with the parameters listed in
Table \ref{fitparams}. In Figure \ref{deltacchar8}, we plot the number of
$\delta_{C}$ computations, per element, for both inter- and intra-cluster
$\eta_{C}$. The 9 and 11 key elements for those two can clearly be seen in
the upper portion of Figure \ref{deltacchar8}; these are dominated by
the $\alpha$ and Fe-peak elements, in addition to Ba, due to their
relative ease of measurement. We stress that these do not necessarily
represent the best chemical tagging discriminators, but rather that
the high ranking elements shown are capable of chemical tagging at the
level depicted in Figure \ref{prob8fig}, based on the data available.

\subsection{Probing chemical dimensionality}
\begin{figure}
\centering
\includegraphics[width=1\linewidth]{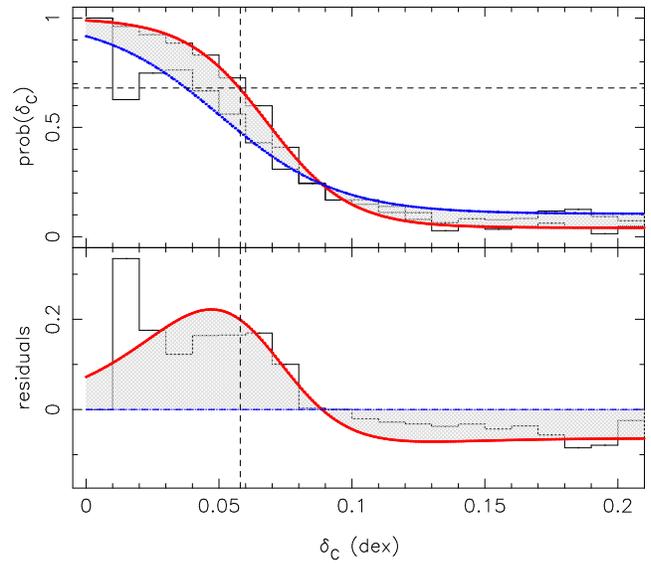} 
\caption{The effects of dimensionality on the probability
function. The top panel shows discrete and fit continuous
probability functions for both the minimum 8 dimensions (upper;red)
and no limit (lower;blue dotted) $\eta_{C}$ distributions. The bottom
panel shows residuals computed from $P_{\delta_C}^8 -
P_{\delta_C}^{all}$ for data (black histogram) and continuous
representations (red line). The strongly positive residuals toward
the intra-cluster portion (left-hand side) of the plot clearly speak
to the necessity of a larger number of dimensions. Cross-hairs are as
in Figure \ref{prob8fig}.
 \label{deltacall}}
\end{figure}

The role of dimensionality on chemical tagging can be tested using the
formulation presented here. By relaxing the requirement that a minimum
number of species be present in each $\delta_{C}$ computation we can
include the entire database summarized in Table \ref{deltactab}, which
results in a minimum of 2 and mean of 5.6 dimensions. Figure
\ref{deltacall} shows the probability functions derived from their
$\eta_{C}$ distributions, including continuous fits, and the residuals
(both discrete and continuous) obtained by the difference
$P_{\delta_C}^8 - P_{\delta_C}^{all}$. It is evident from the excess
in residuals at low $\delta_{C}$ that the reduction in dimensions results
in a significant loss in the ability to perform chemical tagging. Our
ability to explore the same effects for higher dimensions is trumped
by the limited sample of studies with a wide range of element
abundances. The sharp edge seen in both panels of Figure
\ref{deltacchar8} (at rank 8 and 10 in left and right panels,
respectively) illustrates this. The numbers of pairs dramatically
reduce and noise immediately dominates our distributions. For example,
with a minimum of 10 dimensions imposed, the number of inter-cluster
pairs reduces by 80\% and the number of clusters being probed cuts
almost in half, from 27 to 14.

Although it is fairly intuitive at this point to conclude that
increases in chemical dimensionality lead to improvements in the
probability function, it must be noted that improvements, if any,
would not necessarily be linear or predictable. The difficulty in
deriving abundances for some elements (e.g., weak lines, few lines for
EWs, etc.) at a given resolution could yield greater uncertainties and
potential scatter, resulting in little improvement. Indeed, the
limitations on probing higher dimensions in our literature abundance
database likely are a result of this. Furthermore, elements whose
abundance trends are coupled (e.g., see next section) would not
singularly lend a comparable improvement over individual uncoupled
elements.

\subsection{Probing specific elements}
\begin{figure*}
\centering
\includegraphics[width=1\linewidth]{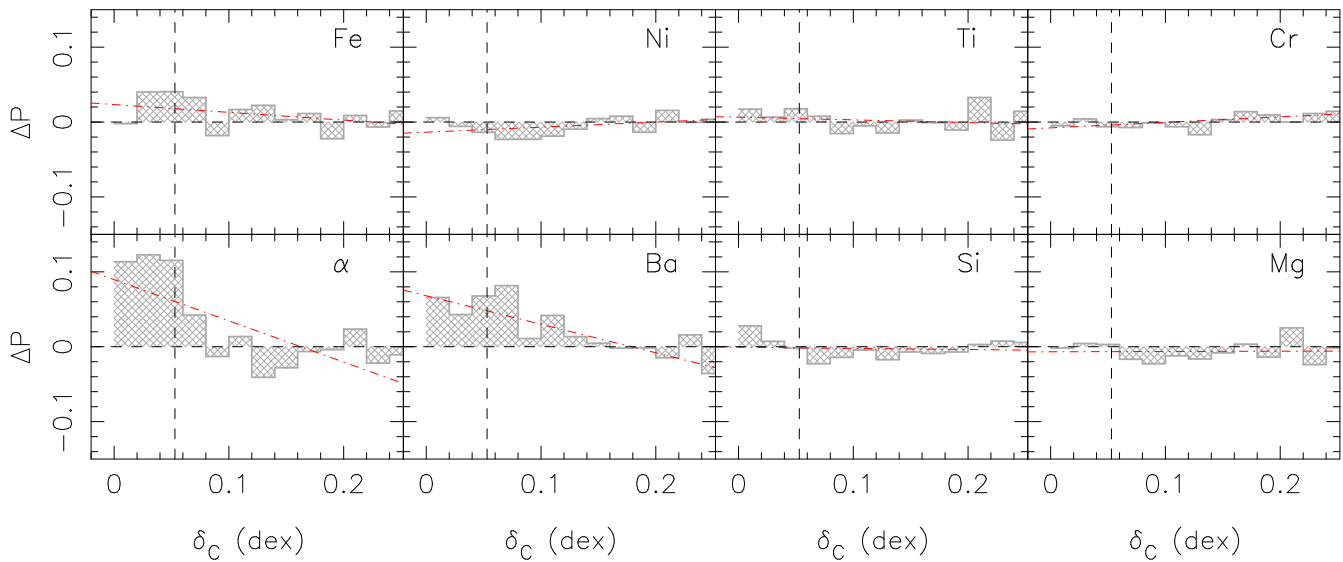} 
\caption{Examining the role of chemical tagging species in the cluster
probability function. Each panel shows the difference residuals
between the function derived in the text and one derived with the
particular element, as indicated, ignored in the $\delta_{C}$ metric
computation. For most of the elements listed, the effect is quite
small and generally linear. The negative slope, and positive
residuals residuals in the low end of the plot without Ba, indicates
its power as a chemical tagging discriminator.
 \label{chemtagel}}
\end{figure*}

It is now clear that a number of chemical dimensions is required to
differentiate clusters in the Galaxy. Each dimension, however, defined
by the efficiency of its nucleosynthesis in the previous generation of
stars, need not contribute evenly to the overall distribution.
Elements whose processes are relatively insensitive to stellar mass,
for instance, will be generated in most cases and the amount may only
depend on a few parameters of the progenitor cloud. We explore the
role of individual species in the probability function by recomputing
the $\delta_{C}$ metric for each pair, \emph{ignoring} that species. The
probability function computed from the resulting distributions is then
compared to that from the originals. Figure \ref{chemtagel} shows the
residuals obtained by subtracting the new function from that depicted
in Figure \ref{prob8fig} (red line) and Table \ref{fitparams}, for a range
of elements with a high ranking (as from Figure \ref{deltacchar8}). Each
panel also plots a linear regression to the residuals in the range
from 0 to 0.25 (beyond this range the functions become extremely noisy
and mostly meaningless), in order to illustrate the general trends. In
these plots, a positive residual can be interpreted to imply that the
element in question contributes above average to cluster
differentiation, i.e., abundance homogeneity is preserved within a
cluster while the abundance dispersion between clusters is increased
for that element. This is particularly true at the low end of the
$\delta_{C}$ spectrum (i.e. intra-cluster probability greater than
approximately the 68\% level) where chemical tagging is possible,
marked by the vertical dashed line in Figure \ref{chemtagel}.

Typically, for the $\alpha$ and Fe peak elements probed, the level of
influence is average to within the noise, at less than about
1-2\%. This is not to say that they are poor chemical tagging elements,
only that they act in concert, providing the robustness and noise
reduction of greater numbers of dimensions. The $\alpha$ panel of Figure
\ref{chemtagel} exemplifies this point, where here the combination of
Na, Mg, Al, Si and Ca acts as a proxy for the $\alpha$ elements. Although
the resultant $\eta_{C}$ must suffer from an increase in noise due to
diminishing numbers of dimensions (the total number of pairs is still
preserved), it is quite clear that the effect on the cluster
probability is significant, at greater than the 10\% level throughout
intra-cluster regime.

Fe and Si seem to be exceptions, contributing closer to the 3-4\% level
on their own. The heavy n-capture element Ba single-handedly alters
the cluster probability by close to 8\%. Ba can be produced both
through the slow and rapid n-capture processes, in low mass AGB
(e.g. \citealt{busso2001}) and core-collapse supernovae
(e.g. \citealt{kratz2007}), respectively, giving a wide range of conditions
owing to the observed scatter in disk field and between Galactic open
clusters \citep{dorazi2009,bensby2005}. The dependence on age found by
\citet{dorazi2009} may also be related to its power as a cluster
discriminator. Different clusters evolving on different time-spans
will exhibit distinct efficiencies in Ba production, which are then
imprinted on the new forming cluster stars. These possibilities imply
that other elements, such as La, Nd and Eu, that form through similar
processes, may lend to a significant improvement to the $P_{\delta_C}$;
however, due to the limited availability of their abundances in the
literature our sample does not contain sufficient numbers to explore
these effects further. We must not discount that non-LTE effects,
which may lead to significant overestimate of Ba abundances in young
stars, could be responsible for some of its perceived power.

There are many factors conspiring to make meaningful comparisons for
other elements difficult, not least of which is the stark drop off in
numbers of pairs for elements beyond those in Figure
\ref{chemtagel}. Provided that there are proper calibrations against
known Galactic open clusters, upcoming large scale abundance surveys,
such as GAL AH, will be highly optimised for these types of
investigations.
\section{Cluster reconstruction}
\label{sec:reconstruction}
\subsection{Mock chemical tagging experiment}
We have already touched on the fact that the overlap between the
intra- and inter-cluster $\eta_{C}$ implies that clustering detection
efficiency can not be 100\%. In order to maximise the confidence with
which we report detections, we must sacrifice potential members in
lieu of including non-members. The actual efficiency which can be
achieved is a function of the area of the overlapping regions between
intra- and inter-cluster $\eta_{C}$ (see Figure \ref{deltacfig}), and the
confidence limit, $P_{lim}$, set on detections (e.g. 68, 90, 98\%). In
order to establish how said overlap affects the recovery of clusters,
and to demonstrate a practical application of the probability function
to a group of cluster candidates, we have performed a blind chemical
tagging experiment using the abundance database discussed herein. A
brief description of the method used follows.

$\delta_{C}$ was computed for each distinct pair of stars, ignoring our
\emph{a-priori} knowledge of cluster membership, thus emulating a true
chemical tagging survey. We then identified all pairs with a
probability $P_{\delta_C}$ greater than $P_{lim}$, where $P_{lim}$ was
evaluated at 68, 85, and 90\%. Starting with the most significant pair
of this subset, stars were added to groups where all pairs in the
resulting group have at most a $\delta_{C}$ corresponding to
$P_{lim}$. After addition to a group, a star was removed from the
subset of significant pairs. When no more stars could be added to the
current group, the next most significant pair was identified and this
process repeated. Grouping was then completed when all pairs of the
subset were exhausted. In this method \emph{all} pairs of members of a
candidate group have at most $\delta_{C}$ corresponding to $P_{lim}$, as
opposed to the group mean. This restriction ensures we identify the
core of a cluster and limits the runaway in linkages that would
otherwise occur. This is not particularly sophisticated, and there may
be more optimal methods such as evolving the cluster mean as members
get added, but it does ensure a strict and invariant interpretation of
$P_{lim}$ for our simple experiment.

\begin{table*}
\begin{center}
\caption{Summary of chemical tagging test experiments
 \label{experimentsum}}
\begin{tabular}{rlccccc}
\hline
 $P_{lim}$  &  $N$  &  Detection Rate  &  Contamination Rate  &  Fragmentation Rate  &  Mean $N_{detect}$  &  Mean efficiency \\
\hline
        68  &   16  &  59\%             &  50\%                 &  25\%                 &  7.50             &  42\%             \\
        85  &   20  &  74\%             &  50\%                 &  50\%                 &  4.85             &  25\%             \\
        90  &    9  &  33\%             &  22\%                 &  57\%                 &  4.67             &  19\%             \\
\end{tabular}
\end{center}
\end{table*}

Table \ref{experimentsum} summarizes the results of our experiment for
$P_{lim}$ of 68, 85 and 90\%. Here, our definition of a cluster requires at
least three members (e.g., a single high significance pair would not
be considered a cluster), and a ``clean'' detection is defined as one
where all recovered group members reside in the same true cluster (but
do not necessarily comprise the whole true cluster). A brief
description of the columns of Table \ref{experimentsum} follows. N gives
the number of clusters detected in total, the Detection Rate compares
N to the number of true clusters in the experiment, which is 27. The
Contamination Rate is a measure of the overall quality of detections
and is the fraction of all detections with members from more than a
single true cluster (e.g. not a clean detection). The Fragmentation
Rate is the fraction of clean detections that are repeats, a high
value of which represents an over-sensitivity to internal cluster
scatter. For example, the 57\% fragmentation rate in $P_{lim}=90$ comes
from the fact that out of the 7 clean detections, 5 were made from
stars of the same Hyades study, making 4 repeats. $N_{detect}$ is simply
the mean number of stars in each group, while the mean efficiency
measures the recovery rate of clean clusters, computed as the number
of detected members divided by the total number of true members.
\subsection{Efficiency and constraints}
\label{sec:efficiency}
Table \ref{experimentsum} illustrates some of the challenges of chemical
tagging and gives an idea of what to expect in terms of efficiency and
contamination. It should be noted, however, that the efficiency
tabulated may not be particularly faithful to a true experiment. Many
of the literature studies have low numbers of stars, while the Hyades
sample of \citet{desilva2006} and \citet{paulson2003} contains 46
members. The stark contrast with little middle ground implies that our
numbers likely represent upper bounds, since few clusters would be
expected to have only several members, although it is unclear how
Galactic dynamical effects, such as disk mixing, spiral scatter, and
radial migration would affect the number of original cluster members
expected to be in the scope of a chemical tagging survey. The
fragmentation rate is also dominated by the Hyades, primarily due to
our requirement of at least three candidate members per detection, and
may represent a lower bound.

It is important to understand that the efficiency tabulated in Table
\ref{experimentsum} does not represent the total efficiency of
recovering clusters, but rather the per detection efficiency of
recovering members of a cluster. The total chemical tagging cluster
detection efficiency is related to at least the detection rate,
contamination rate, fragmentation rate, and mean efficiency
combined. A simple calculation for the $P_{lim}=90$ case gives a total
efficiency of $\sim${}2\% (0.33*(1-0.22)*(1-0.57)*0.19*100\%). This is an
efficiency of placing individual stars in their birth clusters; the
efficiency of detecting any fossil of an existing association would
ignore the mean efficiency column and hence for $P_{lim}=90$ evaluates
to $\sim${}12\%. In a million star survey we might then expect to place
20,000 stars in associations. Given the mean $N_{detect}$ of 4.67, this
would amount to approximately 4300 clusters being recovered in such a
survey, to the 90\% confidence level.

Of course, this is a preliminary number using the literature
abundance database, as are all numbers listed in Table
\ref{experimentsum}; a more realistic approximation would only be
possible using the $P_{\delta_C}$ calculated from a GALAH calibration
program as described in Section \ref{sec:contprobfunc}. It is also
important to note that there is an implicit assumption in this
calculation that all stars were indeed born in clusters similar to
those that we observe today. \citet{bland-hawthorn2010b} suggest, based
on their chemical evolution models, that clusters in the low
metallicity regime, corresponding to early generations of star
formation, are more clumped in chemical space, with a wider range of
abundance variation between clusters. This would certainly result in a
different probability function, leading to higher chemical tagging
efficiencies, since the literature abundance database used here is
restricted to higher metallicities ($>\sim{}-0.35$ dex), due to
the metallicity distribution of Galactic open clusters. However, since
most stars in the Galactic disk are also in the higher metallicity
range, this effect may not be important for most chemical tagging
experiments.
\subsection{Tools for chemical tagging}
\label{sec:tools}
The methods used in these mock experiments are not a recommendation of
how to perform group finding on large data-sets. The requirement of
computing the pair-wise $\delta_{C}$ metric over thousands, or millions,
of stars may not be the most efficient way to identify structure. They
do, however, suggest an algorithm for validating and verifying a group
of potential coeval stars, G, identified within a larger sample of
stars (e.g. a Galactic disk field sample) via any means:

\begin{enumerate}[1.)]
\item  Compute $\delta_{C}$ between all pairs of group G.
\item  Identify the $\delta_{C}$ value, $\delta_{{C}}^{{lim}}$, corresponding to
   desired $P_{lim}$ from $P_{\delta_C}$.
\item  Identify all pairs where $\delta_{C}$ $\le$ $\delta_{C}^{{lim}}$.
\item  All stars making up the pairs from 3.) are set $S_c$
\item  All stars making up the pairs \emph{not} included in 3.) are set $S_{nc}$
\item  G is re-evaluated to be the complement of $S_{nc}$ in $S_c$.
\item  Finally, the cluster detection confidence, $P_{clus}$, is evaluated
   using the mean of $\delta_{C}$ in G.

\end{enumerate}

Aside from ensuring that all pairs in the resultant group have a high
probability of being natal siblings, following the preceding
prescription yields a qualitative assessment of our confidence in the
cluster detection based on a global knowledge of the chemical nature
of Galactic open clusters.

\begin{figure}
\centering
\includegraphics[width=1\linewidth]{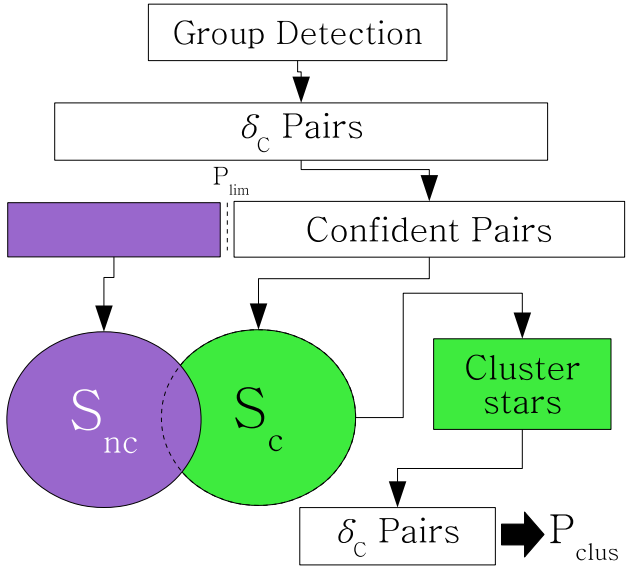} 
\caption{Flow diagram describing the steps outlined in section
\ref{sec:tools} for validation and verification of cluster finding via
chemical tagging. The overlap shown between $S_{nc}$ and $S_c$ represents
the stars that are in both confident and non-confident $\delta_{C}$
pairs. Only those in confident and \emph{not} in non-confident (i.e. the
green region of the $S_c$ circle) comprise the final cluster members.
 \label{flow}}
\end{figure}

Figure \ref{flow} is a graphical representation of the above pipeline,
for clarity. The overlap in the sets $S_{nc}$ and $S_c$ illustrates the
stars that comprise both high and low probability pairs. Only the
stars inside the shaded region of $S_c$ are then included as members in
the final cluster detection. The final number of $\delta_{C}$ pairs shown
at the bottom of the pipeline will be less than the number of
confident pairs (those having $P\ge{}P_{lim}$) shown above, if there
is overlap between $S_{nc}$ and $S_c$ as depicted in this diagram.

This pipeline can be run efficiently, without the need for computing
the $\delta_{C}$ metric over all pairs, with the density based group
finding algorithm, \textbf{EnLink}, described in \citet{sharma2009}. The final
cluster stars derived from the set $S_c$ shown in Figure \ref{flow}, would
result from a $\delta_{C}$ cut in the group detections output by
EnLink. A further investigation of this application, and the overall
use of $P_{\delta_C}$, is forthcoming based on a fully blind chemical
tagging experiment on a homogeneous abundance dataset of several
hundred thin and thick disk field stars.

\section{Summary \& Conclusions}
We have introduced a metric, the $\delta_{C}$ metric, designed to quantify
the chemical difference of a pair of stars in a single context,
regardless of the number or nature of the measured element abundances,
between distinct pairs. Using a database constructed from
high-resolution literature abundance studies of Galactic open
clusters, we have analyzed the relative contributions of $\delta_{C}$
pairs for both intra- and inter-cluster pairs, to arrive at a
probability function, $P_{\delta_C}$, describing the likelihood that
two stars share the same evolutionary origin.

For the specific case of a minimum of 8 chemical dimensions, we have
derived parameters to a functional form of $P_{\delta_C}$, whose 68\%
confidence level (analogous to a 1-$\sigma$ detection) lies at a
$\delta_{C}$ of $\sim${}0.058 dex. This level, in addition to the steep
slope present in $P_{\delta_C}$, informs the accuracy necessary to
distinguish cluster members from polluting stars. We have probed the
effect of variations in the dimensionality of chemical space and the
role of individual elemental species on the probability function. It
is clear that limiting the number of dimensions present severely
reduces the inter-cluster variations, and thus the approximately nine
species prevalent in our analysis are a minimum
recommendation. Unfortunately, owing to the difficulty of measurement,
we could not explore higher dimensions. Of the key players, the $\alpha$
elements (by proxy), Fe and Ba particularly stand out, contributing up
to 10\% of the $P_{\delta_C}$ each.

We have placed these results in the context of chemical tagging, and
shown how they can be used to validate and verify coeval groups
discovered via any means. An important outcome is the ability to
determine and report a confidence of detection, based on our
observational understanding of Galactic structure as opposed to
theoretical models or arbitrary statistical tests. We have utilized
this ability to highlight some of the challenges involved in a true
chemical tagging experiment. It is evident that the reconstruction of
coeval groups of stars, dispersed open clusters, via the technique of
chemical tagging is a difficult, but plausible, venture. Even though
it is impossible to reconstruct all such groups, and seems unlikely
even to discover \emph{most}, the numbers in Table \ref{experimentsum} point
to valuable information that can be inferred about what cannot be
detected, from what can. They also speak to the advantages of a very
large sample of data, given the low efficiencies even at low
confidence levels.

The requirements for internal abundance accuracy of a chemical tagging
survey such as GALAH can be gleaned from the intra-cluster probability
function. However, this function has been derived from the currently
available set of abundance data in the literature which matched our
requirements for resolution, elements, and number of stars, analysed
in several different ways between literature studies. With such a
heterogeneous sample, considerations of systematics effects must not
be ignored, and may result in significant performance loss as
discussed in Sections \ref{sec:systematics} and \ref{sec:probfunc}. The
function derived in this analysis is only the best guess to date and
its use requires understanding the caveats discussed in this text. The
results, however, show that the methods introduced here provide a
powerful tool for conducting and understanding chemical tagging in a
practical sense. To achieve the best possible confidence in cluster
reconstruction, a chemical tagging experiment of large scale should
calibrate against a number of known clusters in order to derive
$P_{\delta_C}$ for the specific analysis methods used therein.

Finally, a pipeline recipe for validation and verification of coeval
groups of stars detected using element abundances only, has been
developed. A logical next step is to use this recipe in a small scale
chemical tagging experiment to further develop these methods in a real
world scenario, and to demonstrate how the tool EnLink can be applied
to this problem. Such an experiment is currently in progress by the
authors on a large homogeneous sample (several hundreds) of stars with
a wide range of high-resolution abundance measurements.
\section*{Acknowledgments}
AWM gratefully acknowledges the Australian Astronomical Observatory for a
PhD top-up scholarship grant which has, and will continue to support
this work. AWM also thanks M. F. Fitzgerald for many stimulating and
useful discussions about the ideas presented here. The authors kindly
thank the anonymous referee for a thorough read of this manuscript and
helpful suggestions which led to clarifications and improvements.
\bibliographystyle{apj}
\bibliography{manuscript}

\begin{thebibliography}{59}
\expandafter\ifx\csname natexlab\endcsname\relax\def\natexlab#1{#1}\fi

\bibitem[{{Barden} {et~al.}(2010){Barden}, {Jones}, {Barnes}, {Heijmans},
  {Heng}, {Knight}, {Orr}, {Smith}, {Churilov}, {Brzeski}, {Waller},
  {Shortridge}, {Horton}, {Mayfield}, {Haynes}, {Haynes}, {Whittard},
  {Goodwin}, {Smedley}, {Saunders}, {Gillingham}, {Penny}, {Farrell}, {Vuong},
  {Heald}, {Lee}, {Muller}, {Freeman}, {Bland-Hawthorn}, {Zucker}, \& {de
  Silva}}]{barden2010}
{Barden}, S.~C., {Jones}, D.~J., {Barnes}, S.~I., {et~al.} 2010, in Society of
  Photo-Optical Instrumentation Engineers (SPIE) Conference Series, Vol. 7735,
  Society of Photo-Optical Instrumentation Engineers (SPIE) Conference Series,
  9--19

\bibitem[{{Batista} \& {Fernandes}(2012)}]{batista2012}
{Batista}, S.~F.~A., \& {Fernandes}, J. 2012, \na, 17, 514

\bibitem[{{Bensby} {et~al.}(2005){Bensby}, {Feltzing}, {Lundstr\"om}, \&
  {Ilyin}}]{bensby2005}
{Bensby}, T., {Feltzing}, S., {Lundstr\"om}, I., \& {Ilyin}, I. 2005, \aap,
  433, 185

\bibitem[{{Bland-Hawthorn} \& {Freeman}(2004)}]{bhf2004}
{Bland-Hawthorn}, J., \& {Freeman}, K.~C. 2004, \pasa, 21, 110

\bibitem[{{Bland-Hawthorn} {et~al.}(2010{\natexlab{a}}){Bland-Hawthorn},
  {Karlsson}, {Sharma}, {Krumholz}, \& {Silk}}]{bland-hawthorn2010b}
{Bland-Hawthorn}, J., {Karlsson}, T., {Sharma}, S., {Krumholz}, M., \& {Silk},
  J. 2010{\natexlab{a}}, \apj, 721, 582

\bibitem[{{Bland-Hawthorn} {et~al.}(2010{\natexlab{b}}){Bland-Hawthorn},
  {Krumholz}, \& {Freeman}}]{bland-hawthorn2010a}
{Bland-Hawthorn}, J., {Krumholz}, M.~R., \& {Freeman}, K. 2010{\natexlab{b}},
  \apj, 713, 166

\bibitem[{{Bragaglia} {et~al.}(2008){Bragaglia}, {Sestito}, {Villanova},
  {Carretta}, {Randich}, \& {Tosi}}]{2008AandA...480...79B}
{Bragaglia}, A., {Sestito}, P., {Villanova}, S., {et~al.} 2008, \aap, 480, 79

\bibitem[{{Bubar} \& {King}(2010)}]{bubar2010}
{Bubar}, E.~J., \& {King}, J.~R. 2010, \aj, 140, 293

\bibitem[{{Busso} {et~al.}(2001){Busso}, {Gallino}, {Lambert}, {Travaglio}, \&
  {Smith}}]{busso2001}
{Busso}, M., {Gallino}, R., {Lambert}, D.~L., {Travaglio}, C., \& {Smith},
  V.~V. 2001, \apj, 557, 802

\bibitem[{{Busso} {et~al.}(1999){Busso}, {Gallino}, \&
  {Wasserburg}}]{busso1999}
{Busso}, M., {Gallino}, R., \& {Wasserburg}, G.~J. 1999, \araa, 37, 239

\bibitem[{{Carrera} \& {Pancino}(2011)}]{carrera2011}
{Carrera}, R., \& {Pancino}, E. 2011, \aap, 535, A30

\bibitem[{{Carretta} {et~al.}(2007){Carretta}, {Bragaglia}, \&
  {Gratton}}]{2007AandA...473..129C}
{Carretta}, E., {Bragaglia}, A., \& {Gratton}, R.~G. 2007, \aap, 473, 129

\bibitem[{{Carretta} {et~al.}(2012){Carretta}, {Bragaglia}, {Gratton},
  {Lucatello}, \& {D'Orazi}}]{carretta2012}
{Carretta}, E., {Bragaglia}, A., {Gratton}, R.~G., {Lucatello}, S., \&
  {D'Orazi}, V. 2012, \apjl, 750, L14

\bibitem[{{Carretta} {et~al.}(2005){Carretta}, {Bragaglia}, {Gratton}, \&
  {Tosi}}]{2005AandA...441..131C}
{Carretta}, E., {Bragaglia}, A., {Gratton}, R.~G., \& {Tosi}, M. 2005, \aap,
  441, 131

\bibitem[{{De Silva} {et~al.}(2007{\natexlab{a}}){De Silva}, {Freeman},
  {Asplund}, {Bland-Hawthorn}, {Bessell}, \& {Collet}}]{desilva2007b}
{De Silva}, G.~M., {Freeman}, K.~C., {Asplund}, M., {et~al.}
  2007{\natexlab{a}}, \aj, 133, 1161

\bibitem[{{De Silva} {et~al.}(2007{\natexlab{b}}){De Silva}, {Freeman},
  {Bland-Hawthorn}, {Asplund}, \& {Bessell}}]{desilva2007a}
{De Silva}, G.~M., {Freeman}, K.~C., {Bland-Hawthorn}, J., {Asplund}, M., \&
  {Bessell}, M.~S. 2007{\natexlab{b}}, \aj, 133, 694

\bibitem[{{De Silva} {et~al.}(2011){De Silva}, {Freeman}, {Bland-Hawthorn},
  {Asplund}, {Williams}, \& {Holmberg}}]{desilva2011}
{De Silva}, G.~M., {Freeman}, K.~C., {Bland-Hawthorn}, J., {et~al.} 2011,
  \mnras, 415, 563

\bibitem[{{De Silva} {et~al.}(2009){De Silva}, {Gibson}, {Lattanzio}, \&
  {Asplund}}]{desilva2009}
{De Silva}, G.~M., {Gibson}, B.~K., {Lattanzio}, J., \& {Asplund}, M. 2009,
  \aap, 500, L25

\bibitem[{{De Silva} {et~al.}(2006){De Silva}, {Sneden}, {Paulson}, {Asplund},
  {Bland-Hawthorn}, {Bessell}, \& {Freeman}}]{desilva2006}
{De Silva}, G.~M., {Sneden}, C., {Paulson}, D.~B., {et~al.} 2006, \aj, 131, 455

\bibitem[{{D'Orazi} {et~al.}(2009){D'Orazi}, {Magrini}, {Randich}, {Galli},
  {Busso}, \& {Sestito}}]{dorazi2009}
{D'Orazi}, V., {Magrini}, L., {Randich}, S., {et~al.} 2009, \apjl, 693, L31

\bibitem[{{D'Orazi} \& {Randich}(2009)}]{2009AandA...501..553D}
{D'Orazi}, V., \& {Randich}, S. 2009, \aap, 501, 553

\bibitem[{{Eggen}(1994)}]{eggen1994}
{Eggen}, O.~J. 1994, in Galactic and Solar System Optical Astrometry, ed. L.~V.
  {Morrison} \& G.~F. {Gilmore}, 191

\bibitem[{{Ford} {et~al.}(2005){Ford}, {Jeffries}, \&
  {Smalley}}]{2005MNRAS.364..272F}
{Ford}, A., {Jeffries}, R.~D., \& {Smalley}, B. 2005, \mnras, 364, 272

\bibitem[{{Freeman} \& {Bland-Hawthorn}(2002)}]{fbh2002}
{Freeman}, K., \& {Bland-Hawthorn}, J. 2002, \araa, 40, 487

\bibitem[{{Friel}(1995)}]{friel1995}
{Friel}, E.~D. 1995, \araa, 33, 381

\bibitem[{{Gebran} \& {Monier}(2008)}]{2008AandA...483..567G}
{Gebran}, M., \& {Monier}, R. 2008, \aap, 483, 567

\bibitem[{{Gebran} {et~al.}(2008){Gebran}, {Monier}, \&
  {Richard}}]{2008AandA...479..189G}
{Gebran}, M., {Monier}, R., \& {Richard}, O. 2008, \aap, 479, 189

\bibitem[{{Gilmore} {et~al.}(2012){Gilmore}, {Randich}, {Asplund}, {Binney},
  {Bonifacio}, {Drew}, {Feltzing}, {Ferguson}, {Jeffries}, {Micela},
  {Negueruela}, {Prusti}, {Rix}, {Vallenari}, {Alfaro}, {Allende-Prieto},
  {Babusiaux}, {Bensby}, {Blomme}, {Bragaglia}, {Flaccomio}, {Fran{\c c}ois},
  {Irwin}, {Koposov}, {Korn}, {Lanzafame}, {Pancino}, {Paunzen},
  {Recio-Blanco}, {Sacco}, {Smiljanic}, {Van Eck}, \& {Walton}}]{gilmore2012}
{Gilmore}, G., {Randich}, S., {Asplund}, M., {et~al.} 2012, The Messenger, 147,
  25

\bibitem[{{Gonzalez} \& {Wallerstein}(2000)}]{2000PASP..112.1081G}
{Gonzalez}, G., \& {Wallerstein}, G. 2000, \pasp, 112, 1081

\bibitem[{{Iben}(1967)}]{iben1967}
{Iben}, J. 1967, \apj, 147, 624

\bibitem[{{Janes} \& {Phelps}(1994)}]{janes1994}
{Janes}, K.~A., \& {Phelps}, R.~L. 1994, \aj, 108, 1773

\bibitem[{{Janes} {et~al.}(1988){Janes}, {Tilley}, \& {Lynga}}]{janes1988}
{Janes}, K.~A., {Tilley}, C., \& {Lynga}, G. 1988, \aj, 95, 771

\bibitem[{{Kaluzny} \& {Rucinski}(1995)}]{kaluzny1995}
{Kaluzny}, J., \& {Rucinski}, S.~M. 1995, \aaps, 114, 1

\bibitem[{{Karakas} \& {Lattanzio}(2007)}]{karakas2007}
{Karakas}, A., \& {Lattanzio}, J.~C. 2007, \pasa, 24, 103

\bibitem[{{Kratz} {et~al.}(2007){Kratz}, {Farouqi}, {Pfeiffer}, {Truran},
  {Sneden}, \& {Cowan}}]{kratz2007}
{Kratz}, K.-L., {Farouqi}, K., {Pfeiffer}, B., {et~al.} 2007, \apj, 662, 39

\bibitem[{{Lada} \& {Lada}(2003)}]{lada2003}
{Lada}, C.~J., \& {Lada}, E.~A. 2003, \araa, 41, 57

\bibitem[{{Magrini} {et~al.}(2010){Magrini}, {Randich}, {Zoccali}, {Jilkova},
  {Carraro}, {Galli}, {Maiorca}, \& {Busso}}]{2010AandA...523A..11M}
{Magrini}, L., {Randich}, S., {Zoccali}, M., {et~al.} 2010, \aap, 523, A11

\bibitem[{{Mitschang} {et~al.}(2012){Mitschang}, {De Silva}, \&
  {Zucker}}]{mitschang2012}
{Mitschang}, A.~W., {De Silva}, G.~M., \& {Zucker}, D.~B. 2012, \mnras, 422,
  3527

\bibitem[{{Montes} {et~al.}(2001){Montes}, {L{\'o}pez-Santiago}, {G{\'a}lvez},
  {Fern{\'a}ndez-Figueroa}, {De Castro}, \& {Cornide}}]{montes2001}
{Montes}, D., {L{\'o}pez-Santiago}, J., {G{\'a}lvez}, M.~C., {et~al.} 2001,
  \mnras, 328, 45

\bibitem[{{Pace} {et~al.}(2008){Pace}, {Pasquini}, \& {Fran{\c
  c}ois}}]{2008AandA...489..403P}
{Pace}, G., {Pasquini}, L., \& {Fran{\c c}ois}, P. 2008, \aap, 489, 403

\bibitem[{{Paulson} {et~al.}(2003){Paulson}, {Sneden}, \&
  {Cochran}}]{paulson2003}
{Paulson}, D.~B., {Sneden}, C., \& {Cochran}, W.~D. 2003, \aj, 125, 3185

\bibitem[{{Pereira} \& {Quireza}(2010)}]{2010IAUS..266..495P}
{Pereira}, C.~B., \& {Quireza}, C. 2010, in IAU Symposium, Vol. 266, IAU
  Symposium, ed. {R.~de Grijs \& J.~R.~D.~L{\'e}pine}, 495--498

\bibitem[{{Phelps} {et~al.}(1994){Phelps}, {Janes}, \&
  {Montgomery}}]{phelps1994}
{Phelps}, R.~L., {Janes}, K.~A., \& {Montgomery}, K.~A. 1994, \aj, 107, 1079

\bibitem[{{Pomp\'eia} {et~al.}(2011){Pomp\'eia}, {Masseron}, {Famaey}, {van
  Eck}, {Jorissen}, {Minchev}, {Siebert}, {Sneden}, {L\'epine}, {Siopis},
  {Gentile}, {Dermine}, {Pasquato}, {van Winckel}, {Waelkens}, {Raskin},
  {Prins}, {Pessemier}, {Hensberge}, {Fr\'emat}, {Dumortier}, \&
  {Bienaym\'e}}]{pompeia2011}
{Pomp\'eia}, L., {Masseron}, T., {Famaey}, B., {et~al.} 2011, \mnras, 415, 1138

\bibitem[{{Randich} {et~al.}(2006){Randich}, {Sestito}, {Primas},
  {Pallavicini}, \& {Pasquini}}]{2006AandA...450..557R}
{Randich}, S., {Sestito}, P., {Primas}, F., {Pallavicini}, R., \& {Pasquini},
  L. 2006, \aap, 450, 557

\bibitem[{{Reddy} {et~al.}(2012){Reddy}, {Giridhar}, \&
  {Lambert}}]{2012MNRAS.419.1350}
{Reddy}, A.~B.~S., {Giridhar}, S., \& {Lambert}, D.~L. 2012, \mnras, 419, 1350

\bibitem[{{Sestito} {et~al.}(2008){Sestito}, {Bragaglia}, {Randich},
  {Pallavicini}, {Andrievsky}, \& {Korotin}}]{2008AandA...488..943S}
{Sestito}, P., {Bragaglia}, A., {Randich}, S., {et~al.} 2008, \aap, 488, 943

\bibitem[{{Sharma} \& {Johnston}(2009)}]{sharma2009}
{Sharma}, S., \& {Johnston}, K.~V. 2009, \apj, 703, 1061

\bibitem[{{Shen} {et~al.}(2005){Shen}, {Jones}, {Lin}, {Liu}, \&
  {Li}}]{2005ApJ...635..608S}
{Shen}, Z.-X., {Jones}, B., {Lin}, D.~N.~C., {Liu}, X.-W., \& {Li}, S.-L. 2005,
  \apj, 635, 608

\bibitem[{{Shu} {et~al.}(1987){Shu}, {Adams}, \& {Lizano}}]{shu1987}
{Shu}, F.~H., {Adams}, F.~C., \& {Lizano}, S. 1987, \araa, 25, 23

\bibitem[{{Smiljanic} {et~al.}(2009){Smiljanic}, {Gauderon}, {North}, {Barbuy},
  {Charbonnel}, \& {Mowlavi}}]{2009AandA...502..267S}
{Smiljanic}, R., {Gauderon}, R., {North}, P., {et~al.} 2009, \aap, 502, 267

\bibitem[{{Soderblom} {et~al.}(2009){Soderblom}, {Laskar}, {Valenti},
  {Stauffer}, \& {Rebull}}]{2009AJ....138.1292S}
{Soderblom}, D.~R., {Laskar}, T., {Valenti}, J.~A., {Stauffer}, J.~R., \&
  {Rebull}, L.~M. 2009, \aj, 138, 1292

\bibitem[{{Tabernero} {et~al.}(2012){Tabernero}, {Montes}, \& {Gonzalez
  Hernandez}}]{tabernero2012}
{Tabernero}, H.~M., {Montes}, D., \& {Gonzalez Hernandez}, J.~I. 2012, \aap, in
  press, ArXiv:1205.4879

\bibitem[{{Tautvai{\v s}iene} {et~al.}(2000){Tautvai{\v s}iene}, {Edvardsson},
  {Tuominen}, \& {Ilyin}}]{2000AandA...360..499T}
{Tautvai{\v s}iene}, G., {Edvardsson}, B., {Tuominen}, I., \& {Ilyin}, I. 2000,
  \aap, 360, 499

\bibitem[{{Ting} {et~al.}(2012){Ting}, {Freeman}, {Kobayashi}, {de Silva}, \&
  {Bland-Hawthorn}}]{ting2012}
{Ting}, Y.~S., {Freeman}, K.~C., {Kobayashi}, C., {de Silva}, G.~M., \&
  {Bland-Hawthorn}, J. 2012, \mnras, 421, 1231

\bibitem[{{Torres} {et~al.}(2008){Torres}, {Quast}, {Melo}, \&
  {Sterzik}}]{torres2008}
{Torres}, C.~A.~O., {Quast}, G.~R., {Melo}, C.~H.~F., \& {Sterzik}, M.~F. 2008,
  {Young Nearby Loose Associations}, ed. B.~{Reipurth}, 757

\bibitem[{{Wallerstein} {et~al.}(1997){Wallerstein}, {Iben}, {Parker},
  {Boesgaard}, {Hale}, {Champagne}, {Barnes}, {K\"appeler}, {Smith}, {Hoffman},
  {Timmes}, {Sneden}, {Boyd}, {Meyer}, \& {Lambert}}]{wallerstein1997}
{Wallerstein}, G., {Iben}, J., {Parker}, P., {et~al.} 1997, Reviews of Modern
  Physics, 69, 995

\bibitem[{{Xin} \& {Deng}(2005)}]{Xin2005}
{Xin}, Y., \& {Deng}, L. 2005, \apj, 619, 824

\bibitem[{{Zuckerman} \& {Song}(2004)}]{zuckerman2004}
{Zuckerman}, B., \& {Song}, I. 2004, \araa, 42, 685

\end{thebibliography}
\end{document}